\documentclass[%
 reprint,
 amsmath,amssymb,
 aps, prx,
]{revtex4-2}

\usepackage{graphicx}
\usepackage{braket}
\usepackage[normalem]{ulem}
\usepackage[utf8]{inputenc}
\usepackage[american]{babel}
\usepackage{graphicx,xcolor,bbold}
\usepackage{braket}
\usepackage{MnSymbol}

\usepackage[breaklinks, pdftex, hyperfootnotes=true, pdfpagelabels, bookmarks, pageanchor]{hyperref}
\usepackage{graphics}
\usepackage{multirow}
\usepackage{mdframed}
\usepackage{tikz,ifthen}
\usepackage{bbold}
\usepackage{tikz-network}
\hypersetup{%
	colorlinks=true, linktocpage=true, pdfstartpage=1, pdfstartview=FitH, pdfborder={0 0 0},%
	breaklinks=true, pdfpagemode=UseNone, pageanchor=true, pdfpagemode=UseOutlines,%
	plainpages=false, bookmarksnumbered, bookmarksopen=true, bookmarksopenlevel=1,%
	hypertexnames=true, pdfhighlight=/O,
	urlcolor=red, linkcolor=red, citecolor=red,
	}
\usepackage{orcidlink}
\usepackage{bbold}
\usepackage{tikz-network}
\usetikzlibrary{patterns,decorations.pathreplacing,calligraphy}
\usetikzlibrary{shapes}
\usetikzlibrary{arrows.meta}
\usetikzlibrary{decorations.pathmorphing}

\definecolor{mycolor2}{rgb}{0.62,0.62,1.}

\begin{document}


\newcommand{\titleinfo}{ Overcoming the entanglement barrier with sampled tensor networks }
\title{\titleinfo}

\author{Stefano Carignano}
\affiliation{Barcelona Supercomputing Center, 08034 Barcelona, Spain}
\email{stefano.carignano@bsc.es}

\author{Guglielmo Lami}
\affiliation{Laboratoire de Physique Th\'eorique et Mod\'elisation, CNRS UMR 8089,
CY Cergy Paris Universit\'e, 95302 Cergy-Pontoise Cedex, France}
\email{guglielmo.lami@cyu.fr}
\email{jacopo.de-nardis@cyu.fr}

\author{Jacopo De Nardis}
\affiliation{Laboratoire de Physique Th\'eorique et Mod\'elisation, CNRS UMR 8089,
CY Cergy Paris Universit\'e, 95302 Cergy-Pontoise Cedex, France}

\author{ Luca Tagliacozzo }
\affiliation{Institute of Fundamental Physics IFF-CSIC, Calle Serrano 113b, Madrid 28006, Spain}
\email{luca.tagliacozzo@iff.csic.es}

\begin{abstract}
The rapid growth of entanglement under unitary time evolution is the primary bottleneck for modern tensor-network techniques—such as Matrix Product States (MPS)—when computing time-dependent expectation values. This {entanglement barrier} restricts classical simulations and, conversely, underpins the quantum advantage anticipated from future devices. Here we show that, for \textit{one-dimensional Hamiltonian dynamics}, the spatio-temporal tensor network encoding the evolved wave function amplitudes can be contracted efficiently along the left–right (spatial) direction. Exploiting this structure, we develop a hybrid Tensor-Network/Monte-Carlo (TN-MC) algorithm that samples the wave function and evaluates expectation values of generic local operators with \emph{computational cost that scales only polynomially in time}. {The accurate contraction of the wave function amplitudes is a consequence of the favorable scaling with time of the generalised temporal entropies. We find that their real part either saturates or, at most, grows logarithmically with time, \textit{revealing new instances of continuous dynamical quantum phase transitions} (DQPTs) which we characterize. Our results therefore show that, when computing expectation values of local operators, the entanglement barrier in one-dimensional Hamiltonian evolution can be bypassed with a TN-MC blend}.
\end{abstract}

\maketitle
\section{Introduction}
Quantum simulation lies at the intersection of theoretical quantum physics and quantum‐technology development. One of its primary aims is to identify tasks that novel quantum devices can perform and that are provably intractable on classical hardware, thereby establishing a clear \emph{quantum advantage}~\cite{Harrow2017,Preskill2018,Lee2023,Haghshenas2025,Arute2019,Kim2023,Begui2024,Morvan2024}. A particularly compelling candidate is the real‐time simulation of chaotic, interacting quantum many‐body systems at large system sizes (or qubit numbers): under generic Hamiltonian dynamics, the bipartite entanglement grows linearly in time, rendering the full wave function impossible to store or manipulate efficiently.

Among the most powerful parameterizations of generic wave functions are \emph{tensor networks} (TNs)~\cite{orus2014,Silvi_2019,ran2020,cirac2021,banuls2023,Collura_2024}. In particular, Matrix Product States (MPS)—the foundation of the celebrated Density Matrix Renormalization Group (DMRG)—are especially well suited for one-dimensional (1D) systems~\cite{affleck1987,white1992,noack1993,ostlund1995}. Although MPS efficiently capture many physically relevant states, such as ground states of 1D local Hamiltonians, under generic real-time evolution the required bond dimension typically grows exponentially with time, due to the rapid increase of bipartite entanglement~\cite{vidal2004,PhysRevA.78.010306,RefPRL2004WhiteFeiguin,daley2004,calabrese_2006,lauchli2008,PhysRevB.104.014301}. Various strategies have been proposed to mitigate this ``entanglement barrier'', including the incorporation of dissipative channels alongside the unitary evolution~\cite{Rakovszky2022,Keyserlingk2022,surace2019a,frias-perez2024,ramos-marimon2025} and unitary changes of basis—e.g., via Clifford circuits—embedded in the MPS ansatz~\cite{PRXQuantum.6.020302,PhysRevB.111.035119,PhysRevLett.124.137701,2407.01692,2502.01872,2408.08249,Qian2024,Fan_2025}. However, these approaches are still insufficiently characterised, and their applicability is only limited to specific cases, such as large temperatures or weak correlations, near-integrable or near-Clifford evolutions and similar cases.

\begin{figure}[t!]
  \centering
  \includegraphics[width=0.98\linewidth]{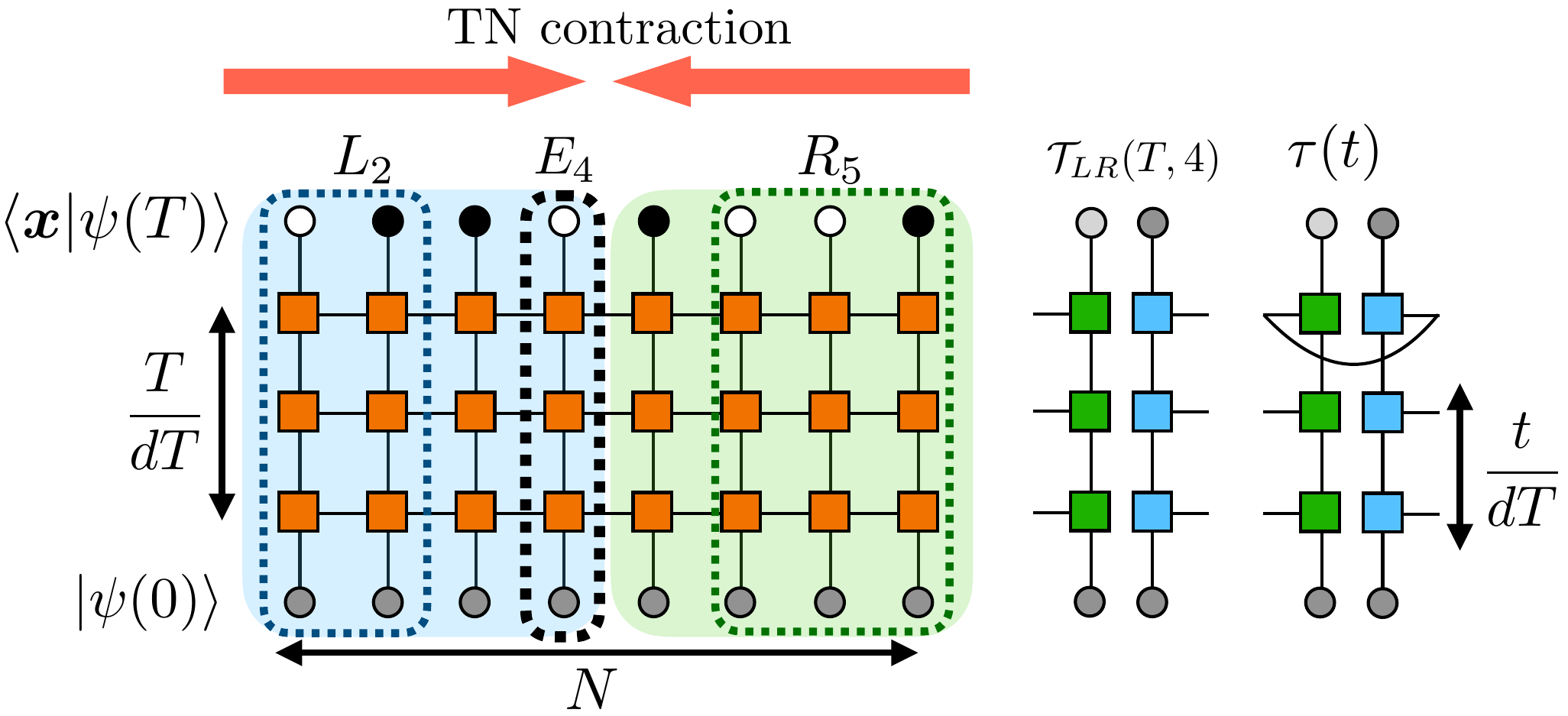}
  \caption{Tensor‐network contraction for the wave function amplitude $\langle\boldsymbol{x}|\psi(T)\rangle$ with respect to a computational‐basis state $|\boldsymbol{x}\rangle$. Starting from the initial product state $|\psi(0)\rangle=|\boldsymbol{x}_0\rangle$, the system undergoes Trotterised evolution with $T/\mathrm{d}T$ steps of duration $\mathrm{d}T$, each implemented by a Matrix Product Operator (MPO). At final time $T$, we project onto $|\boldsymbol{x}\rangle$. Contraction proceeds by building environment tensors $\langle L_i|$ (left of site $i{+}1$) and $|R_i\rangle$ (right of site $i$), and by defining the transition matrix $\mathcal{T}_{LR}(T,i)=|R_i\rangle\langle L_i|/\langle L_i|R_i\rangle$ and its reduced trace $\tau(t)$.}
  \label{fig0:figpaperloschmidt}
\end{figure}

An alternative to conventional time‐directed contraction—which uses an MPS to represent the wave function at each time slice—is to contract along the \emph{spatial} direction. In this approach, one employs \emph{temporal} MPS (tMPS) (i.e. MPS representation of the left/right boundary states whose degrees of freedom have the dimension of the MPO, see Fig. \ref{fig0:figpaperloschmidt}) to encode network slices at fixed spatial locations, and use them to construct temporal environments for the \textit{evolution of local operators }$\langle O(T) \rangle $, requiring therefore a folded Keldysh representation of time evolution ~\cite{banuls2009,muller-hermes2012,Ye2021,hastings2015,RefPRA2015,RefNJP2012,PhysRevX.11.021040,PhysRevB.107.L060305,PhysRevB.107.125103,tirrito2022,carignano2024a,liu2024,cerezo-roquebrun2025,Park2025}.
\begin{figure*}[t!]
  \includegraphics[width=\linewidth]{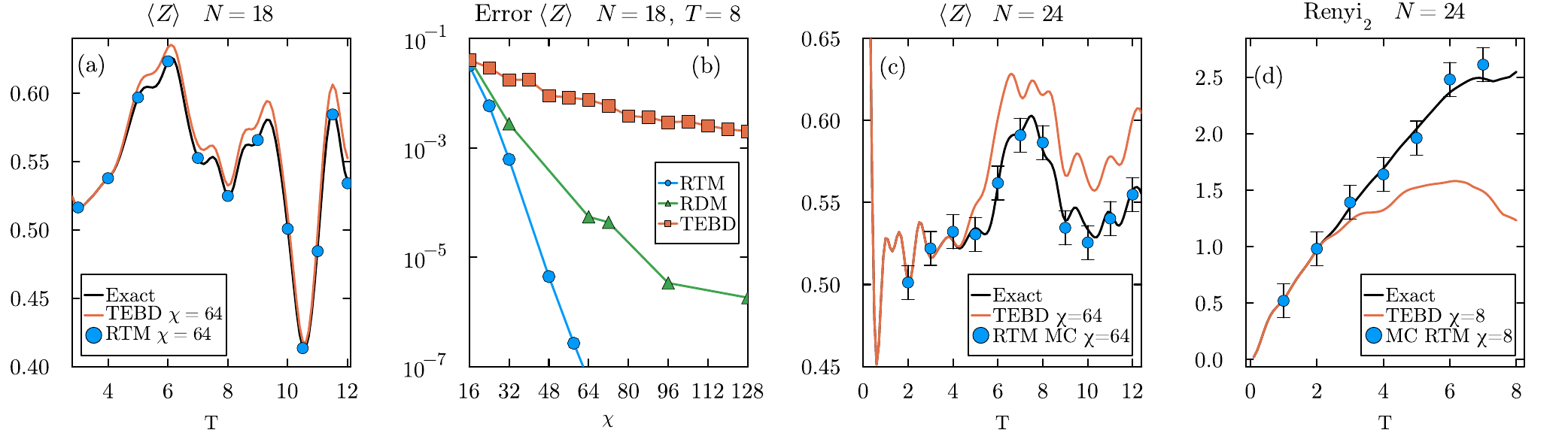}
  \caption{Numerical quench results under the Ising Hamiltonian~\eqref{eq:Hising} with $h_z=-1.05$ and $h_x=0.50$, starting from $\ket{\psi(0)}=\ket{\boldsymbol{0}}$. (a) Time evolution of $\langle Z_{N/2}\rangle$ for $N=18$: blue dots are from our RTM‐based method (max bond dimension $\chi=64$, summing over all $2^N$ configurations), orange squares from TEBD truncated at $\chi=64$, and a dashed line for exact TEBD with $\chi=512$. (b) Absolute error in $\langle Z\rangle$ at $T=8$ for $N=18$, comparing truncated TEBD (orange), individual RDM truncations (green), and our RTM truncation (blue). (c) $\langle Z\rangle$ for $N=24$ via Monte Carlo (MC) sampling ($M=5\times10^3$, $\chi=64$) versus truncated TEBD; error bars denote MC uncertainty. (d) Half‐chain spatial Rényi$_2$ entropy for $N=24$, from MC sampling (blue) versus exact and truncated TEBD.}
  \label{fig:mc}
\end{figure*}
The performance of such algorithms is governed by the \emph{temporal entanglement entropy} of the resulting tMPS. While integrable models or special operators may exhibit modest temporal entanglement, genuinely chaotic Hamiltonians typically manifest \emph{volume‐law} temporal entanglement, reinstating exponential complexity~\cite{PhysRevLett.128.220401,carignano2024,PhysRevX.13.041008}, and therefore limiting the applicability of the method to only specific cases. 

In this work, we propose a different approach: instead of contracting the full TN for expectation values $\langle\psi(T)|\hat{O}|\psi(T)\rangle$, we target the wave function amplitudes 
\begin{equation}\label{eq:waveamp}
   \psi(\boldsymbol{x},T) =  \langle\boldsymbol{x}|\psi(T)\rangle
\end{equation}
with arbitrary bit strings product states $|\boldsymbol{x}\rangle$ (Fig.~\ref{fig0:figpaperloschmidt}). We show that \textit{for generic one-dimensional chaotic evolutions, these amplitudes can be efficiently computed via spatial contraction using our proposed algorithms based on }\emph{temporal reduced transition matrices} (RTMs), a quantity recently introduced in \cite{carignano2024,carignano2024a}, rather than the usual temporal density matrix. Building on this, we devise a Monte Carlo (MC) scheme that samples configurations $\boldsymbol{x}$ to estimate local observables accurately. Our approach concretises tensor‐network Monte Carlo methods—such as TN‐variational MC (TN‐VMC)—a rapidly advancing field~\cite{sandvik2007,ferris2012,ferris2015,PhysRevB.106.L081111,frias-perez2023a,wu2025} which, notably, by compressing amplitudes rather than the wave function itself, has the power to yield volume‐law scaling of entanglement~\cite{liu2024}.

Our key technical insight is that \textit{the generalized temporal entropies of the tMPS encoding wave function amplitudes \eqref{eq:waveamp} grow at most \emph{logarithmically} in time}. Consequently, each amplitude can be approximated with high precision by employing only polynomial computational resources. 

These results extend earlier findings on the Loschmidt echoes~\cite{carignano2024a,gorin2006,serbyn2017,Piroli2017,Andraschko2014,karch2025,2502.01872},
exhibiting logarithmic or area‐law scaling near criticality~\cite{carignano2024a}. We generalise these conclusions to generic bit‐string overlaps: if the real‐time evolution admits analytic continuation to imaginary time without crossing a phase transition, the temporal entropies satisfy an area law; encountering equilibrium or dynamical quantum phase transitions (DQPT)~\cite{heyl2013,heyl2019} induces at most logarithmic divergence, with a universal prefactor characterising each DQPT.

\subsection{Setup and generalized entropies}

We consider $N$ qudits with local dimension $d$, so the Hilbert‐space dimension is $d^N$. A bit string $\boldsymbol{x}=(x_1,\dots,x_N)$, $x_i\in\{0,\dots,d-1\}$, labels computational‐basis states $|\boldsymbol{x}\rangle$. Starting from $|\psi(0)\rangle=|\boldsymbol{x}_0\rangle$, we evolve under a generic (chaotic) Hamiltonian $\hat H$:
\[
  |\psi(T)\rangle = e^{-i\hat H\,T}\,|\psi(0)\rangle.
\]
We focus on the amplitude $\langle\boldsymbol{x}|\psi(T)\rangle$, represented as a spatio‐temporal TN via Trotterisation into $T/\mathrm{d}T$ steps, each of the time steps of the evolution defining an MPO of bond dimension $\chi_{\mathrm{MPO}}$. Cutting vertically at site $i$ exposes $T/\mathrm{d}T$ MPO tensors that propagate \emph{spatially} through network columns $E_i$, viewed as rotated "temporal" MPOs (tMPOs). At each $i$, we define the environment tensors $\langle L_i|$ (left of site $i{+}1$) and $|R_i\rangle$ (right of site $i$), satisfying
\[
  \langle L_{i-1}|\,E_i = \langle L_i|, 
  \quad 
  E_i\,|R_i\rangle = |R_{i-1}\rangle.
\]
Both tensors admit tMPS representations, and moreover they express the wave function amplitudes \eqref{eq:waveamp} as 
\begin{equation}\label{eq:gen_losch}
\langle\boldsymbol{x}|\psi(T)\rangle = \langle L_i|R_i\rangle,
\end{equation}
for any $i \in 1..N$. The left and right environment can be used to define the temporal \emph{transition matrix}
\begin{equation}\label{eq:tm}
  \mathcal{T}_{LR}(T,i)
  = \frac{\ket{R_i}\bra{L_i}}{\braket{L_i|R_i}},
\end{equation}
whose partial traces give the reduced transition matrices (RTMs), which define the \textit{generalized temporal entropies}
\begin{equation}\label{eq:tau}
  \tau(t) = \mathrm{tr}_{T-t}\,\mathcal{T}_{LR}\Bigl(T,\tfrac{N}{2}\Bigr),
  \quad
  S_{LR}(t) = -\mathrm{tr}\bigl[\tau(t)\log\tau(t)\bigr].
\end{equation}
The cost of evaluating $\langle L_i|R_i\rangle$ and therefore the wave function amplitudes, is controlled by the maximum real part of $S_{LR}(t)$, giving the complexity of the transition matrix $ \mathcal{T}_{LR}(T,i)$ at given time $T$. The fact that, as seen below, such entropy grows no faster than logarithmically for generic Hamiltonians and for product boundary states is the key for our TN-MC approach. 
Finally, one can also introduce the temporal 
\emph{density matrix } (either for the $L$ or the $R$ state)
\begin{equation}\label{eq:tm_LL}
  \mathcal{T}_{LL}(T,i)
  = \frac{\ket{L_i}\bra{L_i}}{\braket{L_i|L_i}},
\end{equation}
whose partial trace defines the standard temporal entropy $S_{LL}(t)$, which has been the subject of intense studies in the past years \cite{ippoliti2022,banuls2009,muller-hermes2012,Ye2021,hastings2015,RefPRA2015,RefNJP2012,PhysRevX.11.021040,PhysRevB.107.L060305,PhysRevB.107.125103,tirrito2022}. 
We should stress that the two objects are quite different:
unlike the standard density matrix, 
the transition matrix in general is not hermitian and as such $\tau(t)$ does not have a real positive definite spectrum. As a consequence, the generalised temporal entropies can be complex.
Yet, as we shall argue in the following, they can provide a measure of the complexity associated with their construction via tMPS as discussed in detail below Eq.  \eqref{eq:univ_ent}. Notice also that, again differently from the temporal density matrix, the spectrum of the reduced transition matrices can be efficiently computed only when working with translational invariant (TI) Hamiltonians and initial and final states which are symmetric with respect to reflections around the center. In that case one has $\ket{R_{N/2}} = \bra{L_{N/2}}$, allowing for a diagonalization of $\tau(t)$ with orthogonal rotations.

\subsection{Organization of the manuscript}

In Sect.~\ref{sec:MC} we introduce our TNMC algorithm to sample amplitudes $\langle\boldsymbol{x}|\psi(T)\rangle$ via spatial contraction (Fig.~\ref{fig0:figpaperloschmidt}) at any given time $T$, and show how it can be used to overcome the entanglement barrier arising when computing the expectation values of time-evolved local operator (details on how to construct the transition matrices numerically are given in Appendix~\ref{sec:constructing}). This algorithm constitutes the main result of this work, as it introduces a new efficient way to classically simulate \textit{quantum Hamiltonian dynamics} with sampled TN. 

To motivate its efficiency, in Sec.~\ref{sec:dqpt} we study the slow, logarithmic growth of complexity in the transition matrices given by Hamiltonian evolution, starting from conformal‐field‐theory (CFT) results (also detailed in the Appendix \ref{sec:cft}), and generalising to arbitrary chaotic Hamiltonians. We demonstrate that the logarithmic scaling of generalized temporal entropies is directly related to the presence of dynamical quantum phase transitions in the model. We therefore uncover new universality classes for dynamical phase transitions that appear in the limit of infinite $N$ and $T$.

\section{{Monte Carlo Algorithm for time-dynamics based on Transition Matrices}}\label{sec:MC}

The ability to efficiently represent the amplitudes in Eq.~\eqref{eq:waveamp} allows us to circumvent the entanglement barrier when evaluating the time-dependent expectation value of a local operator~$\hat{O}$,
\[
    \langle O\rangle
    \;=\;
    \langle\psi(T)|\hat{O}|\psi(T)\rangle .
\]
We rewrite this quantity by sampling configurations according to the Born probabilities:
\begin{align}\label{eq:measures}
    \langle O\rangle
    =\sum_{\boldsymbol{x}} p(\boldsymbol{x})\,O_{\mathrm{loc}}(\boldsymbol{x}),
    \qquad
    p(\boldsymbol{x};T)=\bigl|\braket{\boldsymbol{x}|\psi(T)}\bigr|^{2},
\end{align}
where
\begin{equation}\label{eq:oloc}
    O_{\mathrm{loc}}(\boldsymbol{x})
    =\sum_{\boldsymbol{x}'}
      \frac{\braket{\boldsymbol{x}'|\psi(T)}}{\braket{\boldsymbol{x}|\psi(T)}}
      \,\braket{\boldsymbol{x}|\hat{O}|\boldsymbol{x}'},
\end{equation}
is the \emph{local estimator} of~$\hat{O}$~\cite{Becca_Sorella_2017}.  
Equation~\eqref{eq:measures} shows that, if we can  
(i)~sample configurations~$\boldsymbol{x}$ with probability~$p(\boldsymbol{x};T)$ and  
(ii)~evaluate $O_{\mathrm{loc}}(\boldsymbol{x})$ for those samples,  
then $\langle O\rangle$ can be estimated by a simple sample average.  
Task~(ii) is straightforward: because $\hat{O}$ is local, most matrix elements  
$\braket{\boldsymbol{x}|\hat{O}|\boldsymbol{x}'}$ vanish~\cite{Havlicek2023}, and the sum in Eq.~\eqref{eq:oloc} reduces to a handful of terms that differ by only a few local spin flips. For task~(i) one ideally would use exact sampling of the wave function using the Born rule with probability $p(\boldsymbol{x},T)$. However, to apply the Born rule, one needs the knowledge of the exact wave function. Clearly, the full wave function requires an exponential complexity that we want to avoid, as here we claim that what is efficient is the calculation of the amplitudes of the wave function, eq. \eqref{eq:waveamp}. 

Therefore, we employ a standard \textit{Metropolis algorithm}~\cite{Becca_Sorella_2017}:  
We generate a Markov chain $\{\boldsymbol{x}^{(m)}\}_{m=1}^{M}$ by proposing, at each step~$m$, a new configuration $\boldsymbol{x}^{(m+1)}$ obtained from $\boldsymbol{x}^{(m)}$ by flipping a single spin.  
The proposal is accepted with probability  
\[
    p_{\mathrm{acc}}
    =\min \bigl(1,\,
      p(\boldsymbol{x}^{(m+1)};T)/p(\boldsymbol{x}^{(m)};T)\bigr).
\]
A new (decorrelated) sample is recorded each time roughly~$N$ spin flips have been accepted.  
After a suitable thermalisation period, the chain yields configurations distributed according to~$p(\boldsymbol{x};T)$.

The Metropolis updates require only the probability $p(\boldsymbol{x};T)$—equivalently, the amplitude $\langle\boldsymbol{x}|\psi(T)\rangle$—which we obtain from the overlap of the left and right environment tensors via Eq.~\eqref{eq:gen_losch}.  
Taken together, these ingredients constitute an efficient Monte Carlo (MC) scheme for estimating local observables at time~$T$, its efficiency ensured by the bounded growth of the generalized temporal entropies in this setting.

As a proof of concept, we present 
results for system sizes $N=18$ and $N=24$ (the largest sizes for which an exact benchmark is possible using an exact TN wave function with $\chi=2^{N/2}$), demonstrating that our method computes expectation values efficiently.  
Time evolution is governed by the strongly chaotic transverse-field Ising chain, written in Pauli operators $X$,~$Y$,~$Z$ as
\begin{equation}\label{eq:Hising}
    \hat{H}
    =\sum_{j}\bigl(
        X_{j}X_{j+1}
        +h_{z}\,Z_{j}
        +h_{x}\,X_{j}
      \bigr),
\end{equation}
with $h_{z}=-1.05$ and $h_{x}=0.50$. Beyond local observables, we use the MC scheme to estimate the half-chain entanglement entropy of $|\psi(T)\rangle$, defined by the second R\'enyi entropy  
$\mathrm{R\acute{e}nyi}_{2}=-\log \bigl(\mathrm{Tr}\,\rho_{N/2}^{\,2}(T)\bigr)$,  
where $\rho_{N/2}(T)=\mathrm{Tr}_{N/2+1,\dots,N} \bigl[|\psi(T)\rangle\langle\psi(T)|\bigr]$.  
Following Refs.~\cite{buividovich2008,caraglio2008,PhysRevB.81.060411,Gliozzi2010,PhysRevLett.104.157201,PhysRevB.86.235116,Sinibaldi_2023}, we rewrite the purity $\mathrm{Tr}\,\rho_{N/2}^{\,2}$ as the expectation value of a \textsc{Swap} operator acting on two replicas of the system, which we sample in parallel.  
The results are reported in Fig.~\ref{fig:mc}.  
Notably, panel~(b) reveals nearly exponential convergence with respect to the bond dimension by truncating over generalised temporal states.  Furthermore, panel~(d) shows that, in contrast to standard TEBD, our approach accurately captures spatially volume law entangled states even at modest bond dimensions, corroborating recent findings in Ref.~\cite{liu2024}.

In the next section, we discuss the physics behind the favorable algorithmic complexity in the case of continuous Hamiltonian dynamics.\\

\section{{Logarithmic growth of the Generalised entropies and DQPT}}\label{sec:dqpt}

As mentioned above, the complexity of compressing the overlap $\langle L_i| R_i\rangle$ with tMPS for $\langle L_i|$ and $|R_i\rangle$ is governed by the scaling of {generalized temporal entropies}, which are complex-valued quantities that arise from the partial contraction of the transition matrix $\mathcal{T}_{LR}$ defined in Eq. \eqref{eq:tau} (see also Fig. \ref{fig0:figpaperloschmidt}) up to time~$t$~\cite{carignano2024}.  In this section we shall motivate that, in the worst scenario, their real part only grows logarithmically with $T$. These entropies have been studied extensively in holographic field theories~\cite{narayan2023a,narayan2023b,doi2023a,heller2024,milekhin2025a}.  Here, we first recall the results obtained directly from conformal field theory (CFT), which can predict their time scaling, to then move to the treatment of the generic cases.

\begin{figure}[h!]
\centering
\includegraphics[width=0.3\linewidth]{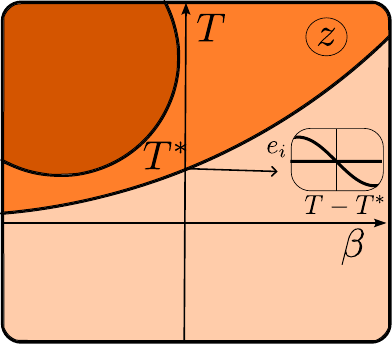}
\caption{  
The wave function amplitude in the complex plane $z=\beta+iT$ exhibits different dynamical phases, illustrated by distinct colours. Real-time dynamics occur along the imaginary axis, whereas the real axis corresponds to equilibrium physics. The physics within a given dynamical phase can be accessed via analytic continuation. Different phases are separated by lines of Fisher zeros; when such lines intersect the time axis, the wave function amplitude becomes non-analytic, signaling the typical DQPTs seen in Loschmidt echoes. In general, these DQPTs are first-order transitions, characterised in the thermodynamic limit by level crossings between the largest and second-largest eigenvalues of the spatial transfer matrix  $E$ (see the inset). 
\label{fig:dqpt}}
\end{figure}

\subsection{Transition Matrices and DQPT}

The central idea to connect the structure of the transition matrix with equilibrium physics is to analytically continue the wave function amplitudes $\langle\boldsymbol{x}|\psi(T)\rangle $ in Eq.~\eqref{eq:gen_losch} across the complex plane by replacing $iT$ with $z=iT+\beta$, as sketched in Fig.~\ref{fig:dqpt}, which also defines an associated free energy $f(\beta) = (N \beta)^{-1} \log | \langle\boldsymbol{x}|\psi(i \beta )\rangle| $. The imaginary axis corresponds to real-time dynamics (the original amplitude), whereas the real axis corresponds to Euclidean time and yields equilibrium matrix elements of thermal states at inverse temperature~$\beta$, evaluated between the same initial and final states.

In the complex plane, multiple dynamical phases may emerge, separated by lines of Fisher zeros, which are vanishing points of the free energy. 
In Fig.~\ref{fig:dqpt}, different dynamical phases are represented by different colours: the entire real axis lies within a single phase, while the imaginary axis spans two of the three phases. When the wave function amplitude reduces to a \textit{Loschmidt echo} (when initial and final state are the same), this phase diagram matches the one underlying dynamical quantum phase transitions (DQPTs)~\cite{heyl2013,heyl2019,PhysRevLett.119.080501,PhysRevLett.121.130603,PhysRevLett.126.040602,PhysRevB.104.075130,PrezGarca2024}. 

Here, we generalise this construction to arbitrary wave function amplitudes and identify DQPTs as the points where Fisher-zero lines cross the imaginary axis; we denote the corresponding critical times by $T^*$ (see Fig.~\ref{fig:dqpt}). Within a single dynamical phase, the analytic continuation is valid, and any point can be reached from any other within the same phase. Crossing a Fisher-zero line, however, induces a breakdown of analytic continuation, as the analytic structure changes discontinuously across the boundary.

\begin{figure*}[ht!]
\centering
\includegraphics[width=0.9\linewidth]{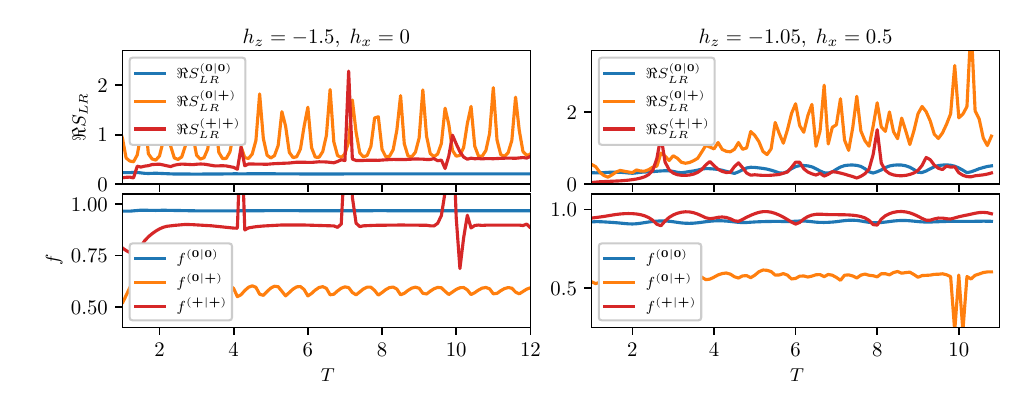}
\caption{
\textit{Left}: Top: generalized temporal entropy for the integrable Ising chain with $h_z = - 1.5$ and $h_x=0$, for the contraction $\langle \boldsymbol{0} (T) | \boldsymbol{0} \rangle $, $\langle  \boldsymbol{+}(T) |  \boldsymbol{+} \rangle $ and  $\langle \boldsymbol{0}(T) |  \boldsymbol{+} \rangle $ in the thermodynamic limit $N=\infty$. Bottom: plot of the free energies in the cases above, using the same time axis to manifest the correspondence between the nonanalyticities in the free energies and the correspondent peaks in the generalized entropy at the DQPT. 
\textit{Right}: same but for the chaotic (non-integrable) Transverse field Ising chain. We remark how in this case the configuration $\boldsymbol{0}|+$ shows DQPT at each times $T$, differently from all the other cases (also in the integrable chain on the left) that instead show isolated or regularly spaced DQPT.     }
\label{fig:dqpt-gen}
\end{figure*}

For simplicity {\textit{we shall now focus on the thermodynamic limit of a translationally invariant system}} and consider situations where both the real and imaginary axes lie in the same phase. In such cases the wave function amplitude are mapped  via the analytic continuation from $iT \to  \beta$ to a boundary partition function on an infinite strip of finite width $\beta$. 

The transfer matrix $E$ becomes a transfer matrix along the strip. The left and right environments correspond to the dominant eigenvectors of the analytically continued $\tilde{E}$, the latter being just the density matrix of the ground state of a system defined on a finite chain of length $v \beta$. Indeed, given that we are considering matrix element of a thermal state, we can safely rotate the strip and, up to trivial rescaling by the sound velocity $v$, we can interpret $v\beta$ as a finite system size. The infinite length of the strip, arising from taking the thermodynamic limit of the original system,  implies we are studying the ground state physics of such a finite chain with boundary conditions dictated by the boundary states.

The generalized transition matrices thus become the reduced density matrices of blocks of such ground state (see also the detail in Appendix \ref{sec:cft}). The operator $\tilde{E}$ is gapped away from criticality and gapless at criticality: when it is gapped, its leading eigenvectors can be efficiently approximated by MPS with constant bond dimension, consistent with the standard area law for entanglement entropy~\cite{hastings2007}. In the gapless scenario instead, the entanglement of a region only grows logarithmically with the size of that region, a properties that ensures that  a bond dimension which increases polynomially with $\beta$ is sufficient to encode these states \cite{verstraete2006}.

By rotating back to real time, the scaling of the standard entropies is inherited by the \emph{real} part of the generalized temporal entropy, which, as a result, grows logarithmically with time exactly at criticality, whereas away from criticality it saturates to a constant (the imaginary part instead always saturates). The pre-factor of the logarithmic growth at mid-chain $t=T/2$ reads:
\begin{equation}\label{eq:univ_ent}
    \Re S_{LR}(T)\;\propto\;\frac{c}{6}\,\log T ,
\end{equation}
where $c$ is the central charge of the CFT~\cite{narayan2023a,narayan2023b,doi2023a}.

These considerations allow us to connect the scaling of the real part of the generalized temporal entropy in time with the complexity of representing $L$ and~$R$ as tMPS when there are no DQPTs. 

However, when the system is no longer confined to a single dynamical phase, as it is the case for a generic chaotic Hamiltonian, the behavior of the wave function amplitude can no longer be inferred  from the equilibrium partition functions. We however proceed to analyze the scaling of the generalized temporal entropies, whose real part we conjecture still dictates the complexity of the reduced transition matrices in analogy with  the case of a single dynamical phase. 

The presence of DQPTs is signaled by  non-analyticities  of the wave function amplitude which in our numerical explorations (and in the literature) always appear in sequences. In the transverse picture, the DQPTs correspond—in the thermodynamic limit—to level crossings in the low-energy spectrum of  $E$ ~\cite{andraschko2014a,banuls2024,carignano2024}, as shown in the inset of Fig.~\ref{fig:dqpt}. In general, at finite $T^*$ these crossings induce first-order dynamical transitions.  In generic Hamiltonian systems, however, we observe that  as $T\to\infty$, the sequence of first-order transitions converges to a second-order transition, characterised by the asymptotic closure of all low-energy gaps in $E$. Such a gapless point has been characterized via the gapless entanglement spectra of the stationary states~\cite{torlai_2014,surace_2020,robertson2022}. In this section, we shall show that the onset of criticality is likewise signalled by the logarithmic growth of the real part of the generalized temporal entropies associated with the dominant eigenvectors $L$ and $R$ of~$E$, mirroring the behaviour of standard quantum critical points. Via Eq.~\eqref{eq:univ_ent}, the pre-factor of this logarithmic growth identifies the universality class of the genuine critical point at $T=\infty$.
A posteriori, our numerical studies thus suggest that the long-time regime of generic chaotic Hamiltonians can again be analytically continued to an equilibrium system, and thus the non-analyticies we encounter at finite time  do not spoil our ability to simulate the long-time dynamics.


\begin{figure*}[t!]
\includegraphics[width=\linewidth]{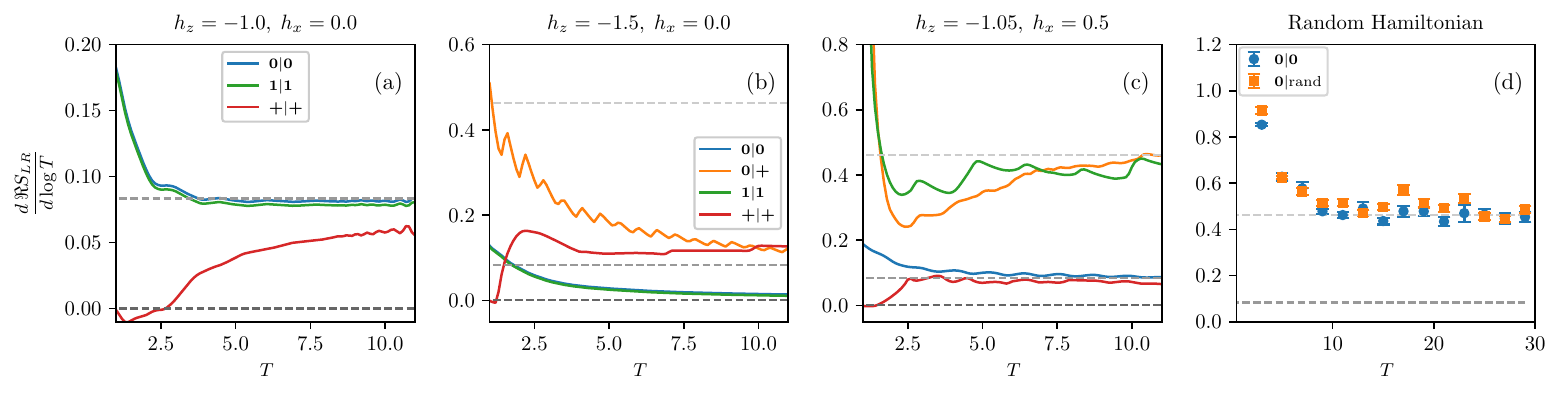}
\caption{Derivatives of the real part of the generalized temporal entropies with respect to $\log T$ for different chains (in the thermodynamic limit $N=\infty$) and for averaged random Hamiltonians (see Fig. \ref{fig02:circuit}) (with $N/2=40$), used to extract the critical charge (data were smoothed, before and after differentiation, to mitigate large oscillations). Plots~(a), (b), and~(c) show data for the Ising chain—integrable and non-integrable—for four different initial/final product states (indicated by $\cdot|\cdot$ for the initial and final states, respectively). Plot~(d) shows data for the case $\boldsymbol{0}|\boldsymbol{0}$ under random-Hamiltonian evolution, see Fig. \ref{fig02:circuit} with $\mathrm{d}T=0.05$, as well as $\boldsymbol{0}|{{\rm rand}}$ where ${\rm rand}$ denotes a random product state (non-translational invariant and invariant but symmetric in respect to reflections around $N/2$), averaged over $200$ realisations with corresponding error bars.}
\label{fig:log_ent}
\end{figure*}

\subsection{Generic Hamiltonians and numerical Results}
 We now present our result for the the {generalized temporal entropies} and the corresponding \textit{free energies} associated with the (cross) Loschmidt echoes 
\[
    f^{(\boldsymbol{x} | \boldsymbol{y}) }(T)=\frac{1}{NT}\,
    \log \Big|\bigl\langle\boldsymbol{x}\bigl|\mathrm{e}^{-i\hat{H}T}\bigr|\boldsymbol{y}\bigr\rangle \Big| ,
    \qquad
    \boldsymbol{x},\boldsymbol{y}\in\bigl\{\boldsymbol{0},\boldsymbol{1},\boldsymbol{+}\bigr\},
\]
where $\boldsymbol{0}$, $\boldsymbol{1}$, and $\boldsymbol{+}$ denote product states with all spins up, down, and along $+x$, respectively, for various choices of the parameters of $\hat{H}$ in Eq.~\eqref{eq:Hising}, and where the notation $\boldsymbol{\cdot}|\boldsymbol{\cdot}$ indicates the initial  $\boldsymbol{x}$ and the final $\boldsymbol{y}$ product state.   We compute these quantities using the algorithm to construct transition matrices presented in Appendix \ref{sec:constructing} that can be adapted to work directly in the thermodynamic limit $N = \infty$.

First, we analyze the relation between DQPTs and generalized entropies: we show that the DQPTs occur not only in Loschmidt echoes, but also in generic matrix elements of the evolution operator, and they are clearly signaled by non-analyticities of the corresponding free energies. Such free energies are represented in the lower panels of Figs. \ref{fig:dqpt-gen} for the integrable and chaotic Ising model. The non-analyticities are aligned in time with corresponding peaks of the real part of the generalized temporal entropies.  We also observe two different scenarios: In the Loschmidt echoes of the integrable chain (left plot) the DQPTs occur at isolated points $T^*$ with well-defined frequencies of occurrence. The case for the Loschmidt echo is well known  \cite{heyl2013}, but also cross-echos display similar sequences of DQPTs.  


By analyzing the scaling of the real part of the generalized temporal entropies, as done in Fig. \ref{fig:log_ent}, we can confirm that, except for the case which shows no DQPT (the $\boldsymbol{0}|\boldsymbol{0}$ configuration in the integrable Hamiltonian with $h_z= -1.5, h_x=0$) the $T\to \infty$ limit of all these other cases leads to a gapless TM at large times and is thus governed by a second-order DQPT in the Ising universality class.

A more complex scenario is observed for the chaotic Hamiltonian dynamics, shown in the right panel in Fig. \ref{fig:dqpt-gen}. For standard echoes, the DQPTs  behave similarly to those in the integrable case and only appear at discrete times $T^ *$, though there seems to be different frequencies dictating their appearance. Also in these cases we observe the emergence of a DQPT in the Ising universality class at $T\to \infty$, as shown in Fig. \ref{fig:log_ent}. The DQPTs for cross-echoes behave instead very differently: They are densely distributed on the time axis, and clearly they occur at a  multitude of frequencies, as manifested by the nowhere continuous shape of the corresponding free energy $f^{\boldsymbol{0}|+}$. Those are the scenarios compatible with the appearance at $T\to \infty$ of a different universality class resembling the strong disorder fixed points,  which we shall describe below.

Moving to the generalized entropy scaling in time, in the \emph{first} panel of Fig.~\ref{fig:log_ent} (a) we verify the CFT predictions by plotting the logarithmic derivative of the generalized temporal entropy for critical $H$ ($h_z=\pm1$, $h_x=0$). Following Ref.~\cite{carignano2024}, we expect this derivative to approach $c/6$ with $c=0.5$, the central charge of the Ising CFT—an expectation confirmed for all boundary-state combinations. The \emph{second} panel of Fig.~\ref{fig:log_ent} (b) shows the same quantity for the case $h_z=-1.5, h_x=0$, which corresponds to a non-critical integrable Hamiltonian. When the real and imaginary axes lie in the same phase, analytic continuation of the CFT predicts an area law for the real part of the generalized temporal entropy, so the logarithmic derivative should vanish, as observed indeed for $\langle\boldsymbol{0}(T)|\boldsymbol{0}\rangle$, where no DQPTs occur~\cite{heyl2013}. For our choice $h_z=-1.5$, DQPTs do occur in $\langle\boldsymbol{+}(T)|\boldsymbol{+}\rangle$; from Refs.~\cite{torlai_2014,surace_2020,robertson2022} we expect that as $T\to\infty$ these DQPTs fall into the Ising universality class, as confirmed by the red curve approaching (albeit fairly slowly) the value $c/6$ with $c=0.5$. The orange curve, $\langle\boldsymbol{+}(T)|\boldsymbol{0}\rangle$, shows a similar trend, likewise indicating an Ising-class DQPT as $T\to\infty$. The \emph{third} panel of Fig.~\ref{fig:log_ent} (c) displays results away from the integrable case (namely with $h_x=0.5$). We again observe convergence to the Ising value for those configurations with isolated DQPT, namely $\langle\boldsymbol{0}(T)|\boldsymbol{0}\rangle$ and $\langle\boldsymbol{+}(T)|\boldsymbol{+}\rangle$ (red and blue curves). The scaling for the cross-matrix echoes is harder to interpret. As discussed above, the non-analyticities of these echoes are dense along the time axis rather than occurring at isolated $T^*$ values. However, we can identify a logarithmic growth with a prefactor trending toward $c_{\mathrm{eff}}=\tfrac16\log D$, with $D=16$ the Hilbert space dimension of four spin-$\tfrac12$’s, indicative of a strongly disordered fixed point \cite{refael2004,refael2009,IGLOI2005,PhysRevLett.90.100601,PhysRevB.74.024427} emerging at $T\to\infty$,  which is quite surprisingly in a clean system.

To probe the non-integrable Hamiltonian cases further, we study the toy circuit of Fig.~\ref{fig02:circuit}, composed of two-body gates $U=\mathrm{e}^{-iA\,\mathrm{d}T}$, where $A$ is a generic non-integrable Hamiltonian sampled from the Gaussian Unitary Ensemble, and chosen to be the same throughout the circuit. Contracted from the left with a boundary tMPS, this circuit (for an evolution time $T$) has $T/\mathrm{d}T$ sites; the number of transverse contractions equals the number of spins. When $\mathrm{d}T\ll1$, the circuit realises a Trotterisation of a generic Hamiltonian, permitting reasoning analogously to the Ising case. The \emph{fourth} panel of Fig.~\ref{fig:log_ent} (d) indeed shows the logarithmic derivative of the real part of the generalized entropy for $\mathrm{d}T=0.05$, revealing a DQPT in the \emph{strong-disorder}   universality class  as $T\to\infty$, evidenced by growth $\Re S_{LR}\sim\tfrac16\log16$. Finally, in all these cases one may also study the standard temporal entanglement, namely the entanglement associated with the temporal density matrix, eq. \eqref{eq:tm_LL}. For this quantity, we cannot make theoretical predictions as  in the previous section, as it does not relate directly to a transfer matrix in the analytical continuation. We observe a close-to-linear growth of the temporal entanglement, in sharp contrast with the generalized entropies (see Fig. \ref{fig02:circuitRandom}). 
\begin{figure}[h!]
\hspace{-8 mm}
\begin{tikzpicture}[baseline=(current bounding box.center), scale=0.65]
\pgfmathsetmacro{\Tt}{5} 
\pgfmathsetmacro{\Nn}{8} 
\pgfmathsetmacro{\Nnn}{\Nn - 1}
\pgfmathsetmacro{\wr}{0.2}
\pgfmathsetmacro{\r}{0.1}
\foreach \t in {1,...,\Tt} {
  \ifodd \t
    \foreach \n in {1,...,\Nnn} {\ifodd\n \draw[thick, black] (\n - 1/2, \t - 1/2) -- (\n + 1/2, \t + 1/2); \draw[thick, black] (\n - 1/2, \t + 1/2) -- (\n + 1/2, \t - 1/2); \draw[line width=\wr mm, fill=mycolor2] (\n - \wr, \t - \wr) rectangle (\n + \wr, \t + \wr); \draw[thick] (\n-\wr/3, \t) -- (\n, \t+\wr/3) -- (\n+\wr/3, \t); \else \fi}
  \else
    \foreach \n in {1,...,\Nnn} {\ifodd\n \else \draw[thick, black] (\n - 1/2, \t - 1/2) -- (\n + 1/2, \t + 1/2); \draw[thick, black] (\n - 1/2, \t + 1/2) -- (\n + 1/2, \t - 1/2); \draw[line width=\wr mm, fill=mycolor2] (\n - \wr, \t - \wr) rectangle (\n + \wr, \t + \wr); \draw[thick] (\n-\wr/3, \t) -- (\n, \t+\wr/3) -- (\n+\wr/3, \t); \fi}
  \fi
}
\foreach \t in {1,...,\Tt} {
  \ifodd \t
  \else
    \draw[thick, black] (1/2, \t - 1/2) -- (0, \t) -- (1/2, \t + 1/2);
    \draw[thick, black] (\Nn-1/2, \t - 1/2) -- (\Nn, \t) -- (\Nn-1/2, \t + 1/2);
  \fi
}
\foreach \n in {1,...,\Nn} {\draw[line width=0.2 mm, fill=white] (\n-1/2, 1/2) circle (\r); \draw[line width=0.2 mm, fill=white] (\n-1/2, \Tt+1/2) circle (\r);}
\draw[thick, dotted] (-0.25,0) -- (-0.25, \Tt+1) -- (\Nn/2+0.3, \Tt+1) -- (\Nn/2+0.3, 0) -- (-0.25, 0);
\node[scale=1.2] at (\Nn/2, \Tt+3/2) {$\langle L |$};
\draw[thick, >=stealth, <->] (-0.5,0) -- (-0.5,\Tt+1);
\node[scale=1.2] at (-1, \Tt/2+1/2) {$\frac{T}{\mathrm{d}T}$};
\draw[thick, >=stealth, <->] (0,-0.25) -- (\Nn,-0.25);
\node[scale=1.2] at (\Nn/2, -0.65) {$N$};
\end{tikzpicture}
\hspace{-1 mm}
\begin{tikzpicture}[baseline=(current bounding box.center), scale=0.45]
\pgfmathsetmacro{\Tt}{5} 
\pgfmathsetmacro{\Nn}{8} 
\pgfmathsetmacro{\Nnn}{\Nn - 1}
\pgfmathsetmacro{\wr}{0.2}
\pgfmathsetmacro{\r}{0.1}
\pgfmathsetmacro{\n}{1}
\pgfmathsetmacro{\t}{1}
\pgfmathsetmacro{\ing}{2}
 \draw[thick, black] (\ing*\n - \ing*1/2, \ing*\t - \ing*1/2) -- (\ing*\n + \ing*1/2, \ing*\t + \ing*1/2); \draw[thick, black] (\ing*\n - \ing*1/2, \ing*\t + \ing*1/2) -- (\ing*\n + \ing*1/2, \ing*\t - \ing*1/2); \draw[line width=\wr mm, fill=mycolor2] (\ing*\n - \ing*\wr, \ing*\t - \ing*\wr) rectangle (\ing*\n + \ing*\wr, \ing*\t + \ing*\wr); \draw[thick] (\ing*\n-\ing*\wr/3, \ing*\t) -- (\ing*\n, \ing*\t+\ing*\wr/3) -- (\ing*\n+\ing*\wr/3, \ing*\t);
\node[scale=1.2] at (\ing*\n- \ing*1 - 0.2, \ing*\t + 0.2 ) {$e^{-i A \mathrm{d}T}=$};
\end{tikzpicture}
\caption{We consider the circuit above, where gates are given by $U = e^{-i A \mathrm{d}T}$, with $A$ a generic GUE matrix. Cusps indicate the direction of unitarity. The resulting state represents the left (or right) environment in the spatial contraction, as in Fig. \ref{fig0:figpaperloschmidt}, where we also use $\ket{R_{N/2}} = \bra{L_{N/2}}$ by imposing left-right symmetry.}
\label{fig02:circuit}
\end{figure}
Remarkably, the continuous time limit is surprisingly much different from the Floquet case, from the point of view of the generalized entropy. Indeed, in a similar setting, one may also consider a generic Floquet circuit with $\mathrm{d}T=1$, where, in accordance with Refs.~\cite{lu2021a,ippoliti2022}, the real part of the generalized temporal entropy and the standard temporal entanglement grows linearly with time (see Fig. \ref{fig02:circuitRandom}). Such a difference between discrete and continuous time evolution was never noticed before, and it will need further clarification in the near future.

\begin{figure}[t!]
\centering
\includegraphics[width=0.9\linewidth]{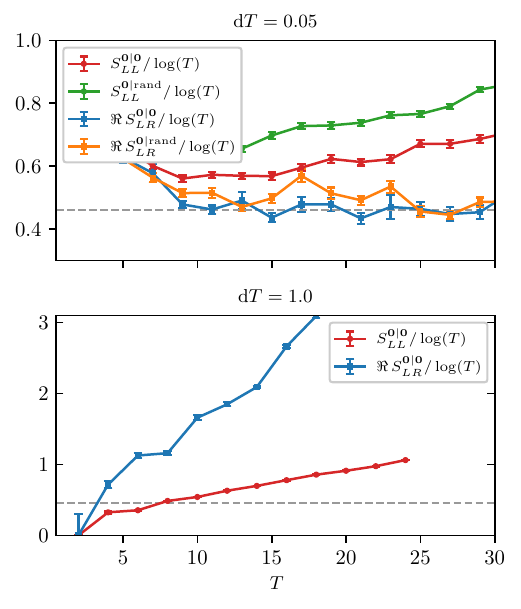}
\caption{ Evolution of temporal entanglement $S_{LL}$ and the generalized temporal entropy $S_{LR}$ for Hamiltonian evolution with $\mathrm{d}T=0.05$ (\textit{Top}) and  Floquet circuits (\textit{Bottom})  with $\mathrm{d}T=1$, divided by the logarithm of time and averaged over 100 different realisation of the circuit Fig. \ref{fig02:circuit} for different initial and final configurations, see also Fig. \ref{fig:log_ent}. The dashed horizontal line indicates the value $1/6 \log 16$.  We observe that the Floquet cases (Bottom) show volume law, namely linear $\sim T$, growth of the generalized temporal entropy and of the temporal entanglement differently from the Hamiltonian case (Top) which shows logarithmic growth of the generalised entropies and linear growth of temporal entropies.   }
\label{fig02:circuitRandom}
\end{figure}

{While the above analysis has focused on the translational invariant setting, the generic wave function amplitudes of Eq. \eqref{eq:gen_losch} are not translational invariant when we project on a generic product state  $\boldsymbol{x}$, in order to compute wave function amplitudes, eq. \eqref{eq:waveamp}. In panel (d) of Fig. \ref{fig:log_ent} we therefore also include the data for a generic Hamiltonian with configuration $\boldsymbol{0}|{\rm rand}$ namely, with the final state given by a random product state, yet, chosen to be symmetric respect to the middle of the chain $i=N/2$ to ensure efficient diagonalisation of the reduced transition matrix. We observe that such generic $\boldsymbol{x}$ at the top boundary of the circuit \ref{fig02:circuitRandom} induces similar results to the TI case, with the worst-case scenario still involving a logarithmic growth in time of $\Re(S_{LR})$. We confirm this claim  also by inspecting the time evolution of the Renyi$_2$ generalised entropy in the chaotic spin chain, averaged over the different states generated in the MC scanning used to produce the results of Fig. \ref{fig:mc}, see Fig. \ref{fig:entropies_mc}. Notice that here the final states are not symmetric and the full diagonalization of $\tau(t)$ is therefore challenging, we focus therefore only on Renyi$_2$ entropies which are simpler to compute using our MPS setup. Analogously to the example of the circuit, we observe the generalized entropies growing logarithmically in time, unlike the temporal entanglement, which instead shows a (quite slow) linear growth. The latter results therefore confirm the \textit{strong advantage of our algorithm based on the transition matrices} for the time evolution of the wave function amplitudes.

\section{{Conclusions}}
We have demonstrated that, thanks to the favorable scaling of {generalized temporal entropies}, a hybrid strategy combining \emph{tensor-network (TN) contraction} with \emph{Monte-Carlo (MC) sampling} can efficiently simulate the time evolution of local operators under generic Hamiltonian dynamics, thereby overcoming the entanglement barrier associated with the evolution of the state or of the operators. We provide  an algorithm based on an efficient finding (in strong analogy with the celebrated DMRG algorithms) the best MPS approximation of the temporal vectors  $| R \rangle, | L \rangle$ by optimizing their relative overlap, which gives the time-evolved wave function amplitudes. We show that the complexity of constructing such tensors only grows polynomially in time and we relate such { logarithmic growth} of the generalized temporal entropies to {dynamical quantum phase transitions (DQPTs)} in the long-time limit, showing that generic quantum systems feature an emergent universal critical point that governs their continuous-time evolution. 

As our algorithm is based on MC sampling of wave function amplitudes, we confirm the recent observation of Liu \emph{et al.}~\cite{liu2024} that encoding wave function amplitudes with TN allows representing states with \textit{spatial} volume law entanglement. However, differently from what claimed in \cite{liu2024}, we find that TN compression of each amplitude is  efficient for continuous- Hamiltonian- time evolution while it remains exponentially complex for a discrete Floquet evolution where the temporal entanglement grows linearly with time, see Fig. \ref{fig02:circuitRandom}. Whether such exponential growth for Floquet one-dimensional dynamics can be eventually curbed remains an important question for the future.  


Several natural extensions now come into view: sampling only states inside the operator’s light cone in the thermodynamic limit, adapting the algorithm to two-dimensional systems, and evaluating finite-temperature correlations within the same framework. In particular, the problem of computing charge fluctuations in one-dimensional chains represents an ideal setting for our approach, which indicates that there is no strict quantum advantage \cite{Rosenberg2024} for such sampling tasks. Finally, by moving beyond known integrable cases, our study reopens the exploration of DQPTs and raises fresh questions about new possible fixed critical points. Furthermore it would be interesting to characterise fixed computational trees, in those scenarios in which temporal entanglement exceeds what can be encoded on boundaries tMPS in order to see how such strategy compares with ours, and if one can still predict efficiently the behavior of local observables even in those cases.\\

\begin{acknowledgments}
 J.D.N. and G.L. are founded by the ERC Starting Grant 101042293 (HEPIQ) and the ANR-22-CPJ1-0021-01. We thank Romain Vasseur, Curt von Keyserlingk, Sarang Gopalakrishnan, Laurens Vanderstraeten, Tibor Rakovszky
 for useful feedbacks.  
LT acknowledges discussions with MC Bañuls, A. Bou Comas, E. Lopez, S. Cerezo and J. Schneider on topics related to this project and the  support from the Proyecto Sinérgico CAM Y2020/TCS-6545 NanoQuCo-CM,
the CSIC Research Platform on Quantum Technologies PTI-001, and from the Grant TED2021-130552B-C22 funded by MCIN/AEI/10.13039/501100011033 and by the ``European Union NextGenerationEU/PRTR'',
and Grant PID2021-127968NB-I00 funded by MCIN/AEI/10.13039/501100011033.
SC acknowledges his AI4S fellowship within the “Generación D”
initiative by Red.es, Ministerio para la Transformación Digital y de la Función Pública, for
talent attraction (C005/24-ED CV1), funded by NextGenerationEU through PRTR.
\end{acknowledgments}

\appendix

\begin{figure}[t!]
\centering
\includegraphics[width=.9\linewidth]{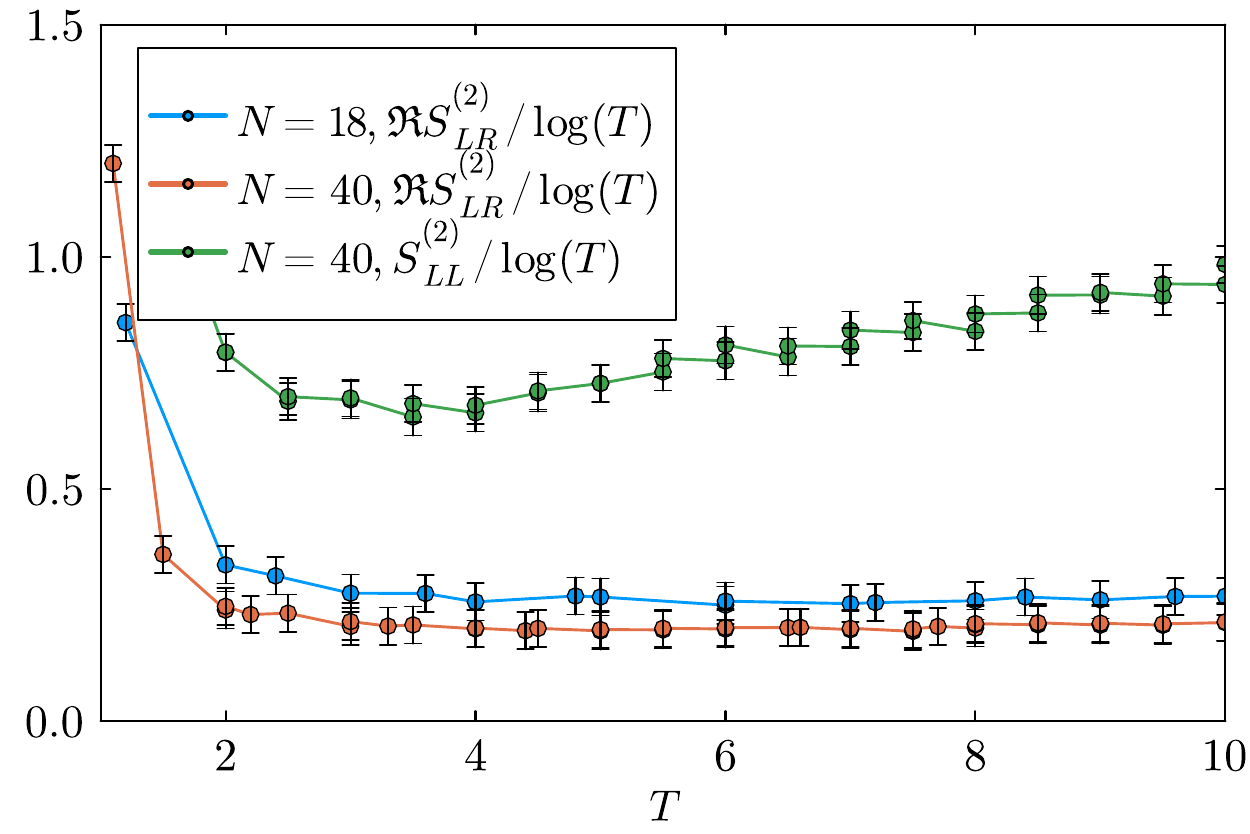}
\caption{Renyi${}_2$ generalised entanglement entropies  and temporal entanglement averaged over typical states that are generated in the MC average to obtain local expectation values in the evolution of the chaotic traverse field Ising chain $h_z=-1.05,h_x=0.5$. We observe a very similar trend to Fig. \ref{fig02:circuitRandom} where the generalised entropies display logarithmic growth, in contrast with the standard temporal entropy that instead grows linearly in time.   
\label{fig:entropies_mc}}
\end{figure}

\section{ {Transitions matrices in CFT}}\label{sec:cft}

In this section, we recall how the CFT strategy introduced in \cite{carignano2024} allows to predict the behavior of the transfer‐matrix spectrum \(E\) and that of the generalized temporal entropies. We focus on the Loschmidt echo of the state \(\boldsymbol{x}\) after an evolution of \(T\) in the thermodynamic limit. The corresponding field‐theory path integral is performed over an infinite strip of width \(T\), whose boundaries are determined by \(\boldsymbol{x}\).

 When interested in characterizing the reduced transition matrices, the associated path integral contains a vertical cut from a point on the boundary to a point in the interior of the strip, of length \(t < T\). Here, \(E\) denotes the transfer matrix along the strip. As usual, we rotate real time to Euclidean time: what used to be time \(T\) now plays the role of \(v\beta\), i.e., the inverse temperature (times the sound velocity). In the Euclidean picture, time corresponds to the vertical direction of the strip (see Fig.~\ref{fig:dqpt}).

For critical Hamiltonians, we can use Conformal Field Theory (CFT) to predict the spectrum of \(E\) \cite{cardy1986,cardy1986a}. This is done by considering Euclidean spacetime as the complex plane parametrized by a coordinate \(\eta\) (note that this is not the same complex coordinate \(z\) used in Fig.~\ref{fig:dqpt}). Since in two dimensions conformal symmetry is equivalent to invariance under analytic changes of coordinates, we perform the conformal transformation that sends the upper half‐plane to the strip. The map is given by
\begin{equation}
    \omega \;=\; f(\eta) \;=\; \frac{\beta\,v}{\pi}\,\log(\eta)\,.
    \label{eq:cf_map1}
\end{equation}
Under this map, a dilation in the \(\eta\)-plane becomes a translation along the strip. Consequently, \textit{the spectrum of \(E\) is dictated by the Virasoro generator \(L_0\)}, which is diagonal on the basis of the scaling fields. One can then predict its spectrum \(\{e_i\}\) \cite{cardy1986a} as 
\begin{equation}
    e_i \;=\; \frac{\kappa(c)}{\beta v} \;+\; \frac{\pi}{v\,\beta}\,\Delta_i \,,
    \label{eq:spect_E}
\end{equation}
where the \(\Delta_i\) are the boundary critical exponents of the theory, selected by the conformally invariant boundary states that encode \(\boldsymbol{x}\) \cite{calabrese_2004,cardy1986a}, $\kappa(c)$ is a universal constant  which depends on the central charge of CFT $c$ and the conformal map. For the case of a strip, $\kappa(c)=\frac{c \pi}{24}$.
Note that for any finite \(v\beta\), the transfer matrix is \textit{gapped}, and that the algebraic closure of this gap as \(\beta \to \infty\) guarantees that its eigenvectors can be approximated by MPS with polynomial resources \cite{verstraete2006}.

We now turn to the generalized temporal entropies and the reduced transition matrices. The idea is again to use conformal invariance to map the strip with an open cut to an annulus. Following \cite{cardy_2016}, the required conformal transformation is
\begin{equation}
    \omega \;=\; f(\eta) \;=\; \log \Biggl[\frac{\sin\bigl(\pi(\eta - \tau)/2\beta\bigr)}
    {\cos\bigl(\pi(\eta + \tau)/2\beta\bigr)}\Biggr]\,.
    \label{eq:cf_map}
\end{equation}
The width \(W\) of the annulus is then given by
\begin{equation}
    W(\beta, \tau) =f\Bigl(\tfrac{\beta}{2}\Bigr) - f\Bigl(\tfrac{\beta}{2} - \tau + \delta\tau\Bigr)
    = \log \Biggl[\frac{2\,\beta\,v}{\pi\,\delta\tau}\,\sin \Bigl(\frac{\pi\,\tau}{\beta}\Bigr)\Biggr]\,,
    \label{eq:W}
\end{equation}
as in \cite{cardy_2016}. Using this map and the connection between Rényi entropies and the correlation function of twist fields \cite{calabrese_2004}, we obtain
\begin{equation}
    S_n \;=\; \frac{c}{12}\Bigl(1 + \tfrac{1}{n}\Bigr)\,W(\beta,\tau)\,.
\end{equation}

Up to this point, these results are standard. Following the strategy of \cite{carignano2024}, and noting that for a \textit{critical Hamiltonian there are no DQPTs} that prevent analytic continuation, we analytically continue back to real time via
\begin{equation}
    \beta \;\longrightarrow\; i\,T \;+\; 2\,\beta_0\,, 
    \qquad
    \tau \;\longrightarrow\; i\,t,
    \label{eq:ana_cont}
\end{equation}
where \(\beta_0\) can be considered as a UV cutoff. Under this continuation, the entanglement entropies become generalized temporal entropies:
\begin{align}
    S_n 
    &= \frac{c}{12}\Bigl(1+\tfrac{1}{n}\Bigr)\,W(T,t)
    \nonumber\\
    &= \frac{c}{12}\Bigl(1 + \tfrac{1}{n}\Bigr)\,
    \log\Biggl[\frac{2\,\bigl(i\,T + 2\,\beta_0\bigr)\,v}{\pi\,\delta t}\,
    \sin\!\Bigl(\tfrac{i\,\pi\,t}{\,i\,T + 2\,\beta_0\,}\Bigr)\Biggr]\,.
\end{align}
In the limit \(\beta_0 \to 0\), this simplifies to
\begin{align}
    S_n 
    &= \frac{c}{12}\Bigl(1 + \tfrac{1}{n}\Bigr)\,
    \log\Bigl[\tfrac{2\,T\,v}{\pi\,\delta t}\,\sin\bigl(\tfrac{\pi\,t}{T}\bigr)\Bigr]
    \;+\; i\,\frac{c\,\pi}{24}\Bigl(1 + \tfrac{1}{n}\Bigr)\,.
\end{align}
The imaginary part of the generalized entropy is constant, while the real part inherits its time dependence from the usual ground‐state entanglement entropy scaling. In the main text, we argue that the scaling also in the presence of DQPT remains at most logarithmic, even if the prefactor can be strongly influenced by their presence.

Finally, by applying the same analytic continuation to the prediction for the transfer‐matrix spectrum \cite{carignano2024}, one can also obtain that the (complex) spectrum of the reduced transition matrices:
\begin{equation}
    e_i \;=\; i \Bigl(\tfrac{\kappa}{T\,v} \;+\; \tfrac{\pi\,\Delta_i}{T\,v}\Bigr)\Bigl(1 + i\,\tfrac{2\,\beta_0}{T}\Bigr)\,.
    \label{eq:TM_spectrum}
\end{equation}
Hence, the gaps of the Euclidean transfer matrix are mapped to gaps in the imaginary parts of the real‐time eigenvalues of \(E\).





\begin{figure}[t!]
\centering
\includegraphics[width=.95\linewidth]{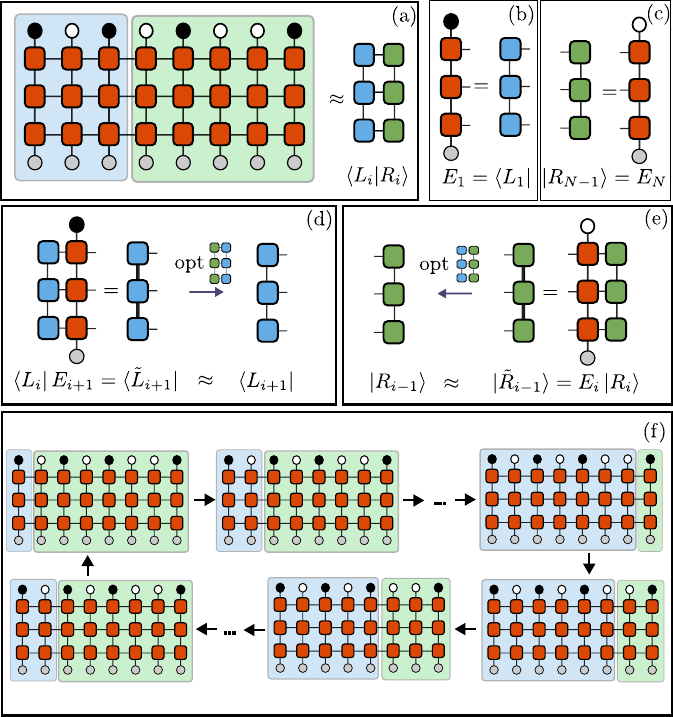}
\caption{Sketch of the Tensor Network methods for efficiently computing amplitudes $\braket{\boldsymbol{x} | \psi(T)}$. (a) For each site $i$ we define a left environment $\bra{L_i}$ and a right environment $\ket{R_{i}}$, given by the transverse contraction of the network for all sites $j\leq i$ and $j>i$, respectively, and represent them using tMPS. The whole amplitude is given by the overlap $\braket{L_i| R_i}$.
(b)-(e) We initialize the environments starting from the edges and progressively applying columns of the network, truncating over reduced transition matrices. (f) We iterate the procedure in a sweeping fashion until the algorithm converges and all $\braket{L_i|R_i}$ give the same result within a desired precision. \label{fig:method}}
\end{figure}

\section{{Constructing transition matrices with Tensor Newtorks}}\label{sec:constructing}

The transverse contraction of the network associated with our amplitudes requires the calculation of left and right tMPS, which can be built iteratively starting from the edges of the system. 
For a finite system, the idea is to define $N-1$ distinct contractions, 
each associated with a different position $i$ of a cut separating the tensor network into a left environment $\bra{L_i}$ and a right environment $\ket{R_{i}}$, given by the transverse contraction of the network for all sites $j\leq i$ and $j>i$, respectively. 
Each overlap $\braket{L_i|R_i}$ then provides a computation of the same amplitude $\braket{\boldsymbol{x}|\psi(T)}$.

Our method is schematically described in Fig.~\ref{fig:method}.
Our goal is to iteratively construct tMPS approximations of each environment. We begin by initializing the boundaries as the external environments $\bra{L_1}$ and $\ket{R_{N-1}}$ corresponding to the first and last columns of the TN. We then evolve the boundary environments in space by applying  columns of tensors that define spatial transfer matrices ${E}_i$.
At each step, the bond dimension of the new environment grows, so that truncation is required.

A common recipe employed in the literature involves truncating the left and right vectors individually on their reduced density matrices (RDM), using standard MPS techniques \cite{banuls2009}. As discussed in the main text, here we will instead
follow the approach introduced in \cite{carignano2024} and focus on building a low-rank approximation of the reduced transition matrices $\tau(t,i)$, obtained by partially tracing $\mathcal{T}_{LR}(i)$ over the final $T-t$ time layers: $\tau(t,i) = \textrm{tr}_{T-t}[\mathcal{T}_{LR}(i)]$.
The optimal tMPS is consequently defined as the one that projects onto the dominant singular values of each reduced transition matrix (For details regarding the appropriate gauge fixing prior to truncation, see \cite{carignano2024}).
Specifically, at each step we obtain a new 
 $\bra{L_i} {E}_{i+1} = \bra{\tilde{L}_{i+1}}$, which after truncation gives the next $\bra{{L}_{i+1}}$, and in the same way we compute ${E}_i \ket{R_i} = \ket{\tilde{R}_{i-1}}$, which after truncation gives $\ket{{R}_{i-1}}$.

Since each environment $\bra{L_i}$ depends on the $\bra{L_{j < i}}$ and, after truncation, also on the corresponding $\ket{R_i}$ (same goes the other way around), the optimization of the environments with this method requires a series of sweeps across the chain, which we perform first from left to right, then from right to left. 
Each iteration yields improved estimates for all $\bra{L_i}$ and $\ket{R_i}$, which are then used to refine the truncation at the next step. 
We perform several such sweeps and monitor convergence by checking that the $N-1$ distinct overlaps $\braket{L_i|R_i}$ yield consistent approximations of the amplitude.
For the first sweep, since we start from the edges towards the center, we can initialize the left environments optimizing their overlap with the closest initialized right one (eg. $\bra{L_2}$ with $\ket{R_{N-2}}$ for the first iteration), or employ a truncation based on RDM with a fixed small bond dimension. These first approximations are quickly improved over subsequent sweeps.  
In practice, we observe that all $L$ and $R$ environments converge after only a few sweeps.

{Finally, we remark that in Ref. \cite{liu2024}, the authors discuss the relevance of designing a consistent contraction method for TN amplitudes. This requires the choice of a computational tree whose structure should be maintained fixed for all amplitudes. This is not what we are doing here, since our computational tree is flexible  given that it mirrors the tMPS truncation which is based on reduced transition matrices computed at different points in space and time. A fixed computational tree would require to use a truncation scheme based on the left and right tMPS taken separately, something that we find suboptimal in this setting, as discussed in the main text.}

\bibliography{bib}
\bibliographystyle{apsrev4-2}

\setcounter{section}{0}
\setcounter{secnumdepth}{2}

\end{document}